\documentclass[preprints,article,accept,moreauthors,dvi2pdf]{mdpi}
\firstpage{1} 
\pubvolume{xx}
\issuenum{1}
\articlenumber{5}
\pubyear{2020}
\copyrightyear{2020}
\history{Received: date; Accepted: date; Published: date}

\pdfoutput=1

\usepackage{graphicx}
\usepackage{dcolumn}
\usepackage{bm}
\usepackage{hyperref}
\usepackage{braket}
\usepackage[title]{appendix}

\Title{A Partial Information Decomposition Based on Causal Tensors}

\Author{David Sigtermans $^{1 *}$} 

\AuthorNames{David Sigtermans}

\address{%
$^{1}$ \quad ASML}

\corres{Correspondence: david.sigtermans@asml.com}

\abstract{We propose a partial information decomposition based on the newly introduced framework of causal tensors, i.e., multilinear stochastic maps that transform source data into destination data. This framework enables us to express an indirect association in terms of the constituting, direct associations. This is not possible when using average measures like mutual information or transfer entropy. From this, an intuitive definition of redundant and unique information arises. The proposed redundancy satisfies the three axioms stated by introduced by Williams and Beer. The symmetry and self-redundancy properties follow directly from our definition. The Data Processing Inequality ensures that the monotonicity axiom is satisfied. Additional, two other proposed axioms are satisfied: the identity property, and the left monotonicity axiom. Because causal tensors can describe both mutual information as transfer entropy, the proposed partial information decomposition applies to both measures. Results show that the decomposition closely resembles the decomposition of another approach that expresses associations in terms of mutual information a posteriori. It is furthermore demonstrated that negative contributions can arise when our assumptions about completeness of the data set, or what should be included as a source, are incorrect.}


\begin{document}

\section{Introduction}
To understand and therefore to \emph{diagnose} changes in the behavior of a specific component of a complex system, we need to explain the behavior of this specific component in terms of the behavior of the other components in that complex system, the sources, and combination of these components, i.e., sets of sources. This interconnectedness of behaviors can be described with information theory \cite{Shannon}. The theory reflects the idea that meaningful communication can only exist if there is an association between the data sent, the source data, and the  data received, the destination data. This association, expressed in mutual information (MI), is modeled as the result of transmission of data over a (conceptual) communication channel. Information theory also gives rise to a succinct description of the interrelations in a multivariate complex system as directed graphs: the ``nodes'', or vertices, are the sources and/or destinations. A directed edge represents unidirectional communication channels between a source and a destination. In the case that the communication channel can not be used to transmit any information, the associated directed edge does not exist. This can be illustrated using a system consisting of three variables: two random binary sources, $X$ and $Y$, and the logical \textsc{exclusive or} combination $X\oplus Y$ as the destination. Because each of the two sources do not share any information with $X\oplus Y$, nor with each other \cite{XOR}, there are no edges between the vertices. This example also illustrates a fundamental aspect of multivariate systems: \emph{``the whole becomes not only more than but very different from the sum of the parts''} \cite{MoreIsDifferent}. The join of the two sources fully determines $X\oplus Y$ \cite{XOR}, i.e., they interact in a \emph{synergistic} fashion. To capture \emph{synergistic} effects that can arise when joining sources, bivariate mutual information is clearly insufficient. Multivariate mutual information measures like the widely used ``Interaction Information'' have been defined  \cite{InteractionInformation}. Unfortunately, this measure is counterintuitive because it can have negative values. An alternative approach is the ``partial information decomposition'' (PID) of Williams and Beer \cite{PID}. The total information shared between the target and the rest of the network is broken down into three non-negative information contributions: \emph{unique}, \emph{redundant}, and \emph{synergistic} information. Unique information is information contributed by one and only one specific source, redundant information is information that is shared between sources, and synergistic information is information that only results from sets of sources, i.e., joined sources. A foundational notion in the PID is that redundancy between a source and any superset, that is, any set of sources containing that source, is equal to the information of that source. This results in an ordering relation between the collections, guaranteeing non-negativity. The ordering relation is captured in the ``redundancy lattice''. While the lattice is widely accepted \cite{InformationLoss}, the proposed redundancy term, $I_{min}$ is not without issues which led to proposals for an (additional) axiomatic underpinning, see for example \cite{LeftChainRule, Identity}. Additionally, different definitions are proposed for the non-negative components \cite{Editorial}.

The main idea behind the redundancy and unique information definitions proposed in this article, is that the underlying issue with respect to a workable definition for redundancy, is the search for an \emph{a priori} expression of associations in terms of mutual information. Notable exceptions are the pointwise approaches of \cite{Ince}, and Finn and Lizier \cite{PPID, GenMeasure}. However, no exact expression exists for ``mediated'', or indirect associations until recently. For example, consider the fully connected system comprising three random variables, say, $X$, $Y$, and $Z$. The indirect association between $X$ and $Z$ resulting from the cascade of associations $X\! \rightarrow\! Y\! \rightarrow\! Z$, cannot be expressed in terms of the associations $X\! \rightarrow\! Y$ and $Y\! \rightarrow\! Z$. The earlier mentioned Interaction Information can be considered the result of such an endeavor. 

When we interpret an association as a communication channel, which is fully characterized by a transition probability matrix \cite{ThomasCover}, linear algebra allows us to express a communication channel resulting from a cascade in terms of the communication channels comprising the cascade. The resulting communication channel gives rise to an association that can also be expressed in mutual information. This \emph{a posteriori} expression of associations in terms of mutual information is used to define unique and redundant information, resulting in a partial information decomposition. Interestingly, although the proposed definitions were not designed to meet proposed axioms, the resulting measures satisfy the three axioms postulated by Williams and Beer, the identity property postulated by Harder \cite{Identity}, and the left monotonicity axiom postulated by Bertschinger et al. \cite{LeftChainRule}. For a system comprising three variables, the ordering relations, as captured in the redundancy lattice, result from the application of the Data Processing Inequality \cite{ThomasCover}.  

The structure of this article is as follows. In the Section \ref{ss:CausalTensors}, causal tensors, a generalization of a transition probability matrix, and the related ``transmission paths'' are introduced. A definition of redundant and unique information is proposed in terms of ``transmission paths'' in Section \ref{ss:PID}. The proposed redundancy meets the three axioms postulated in \cite{PID} and the earlier mentioned additional axioms. The lattice elements of the redundancy lattice correspond to transmission paths: the ordering relation, and therefor the redundancy lattice, now results from the Data Processing Inequality \cite{ThomasCover}. In Section \ref{s:Results}, the proposed PID is applied to some standard systems. In this section a peculiar result is demonstrated. The proposed definitions can result in a negative synergistic contribution \emph{if a variable is unobserved}.
\section{Materials and Methods} \label{s:MandM}
The proposed approach relies on a novel framework inspired by transfer entropy, a conditional mutual information \cite{Schreiber}. In contrast with mutual information, transfer entropy is capable to distinguish a source from a destination, i.e., structures inferred with transfer entropy are expressed, in general, as directed graphs.
\subsection{Causal tensors} \label{ss:CausalTensors}
In information theory, the data are realizations of random variables representing stationary ergodic processes \cite{Shannon}. Because the data comprise elements from a finite alphabet, we can describe an outcome either using the value, i.e., alphabet symbol, or using the index, assuming a fixed, for example, a lexicographic order of the alphabet elements. In this article, we use the latter. The communication channel transforms the probability mass function (pmf) of the source data in the pmf of the destination data via a linear mapping. With $p^j$ representing the $j^{th}$ element of the destination pmf, and $p^i$ representing the $i^{th}$ element of the source pmf, the relation between source and destination is given by

\begin{equation*}
	p^j = \sum_i p^i A^j_i. 
\end{equation*}

The elements of the (row stochastic) tensor $\mathcal{A}$, i.e., $A^j_i$, represent the transition probabilities $A^j_i=p(j^{th}\; destination\; symbol\vert i^{th}\; source\; symbol)$. A communication channel is the conceptual implementation of the Law of Total Probability \cite{LoTP}. This immediately implies that the source pmf can be reconstructed from the destination pmf: $p^i = \sum_j p^j A^{\ddagger i}_j$. The ``$\ddagger$'' indicates that the source pmf is reconstructed. This is relevant in case a source is distinguishable from a destination, e.g., when transfer entropy is used. The mutual information between the source data, generated by process $X$, and the destination data, generated by the process $Y$, is expressed as

\begin{equation} \label{eq:MI}
	I(X,Y) = \sum_{i,j} p^{ij}\log_2 \left[ \frac{A^j_i}{p^j} \right].
\end{equation}

Equation \ref{eq:MI} is equivalent to standard expression for mutual information \cite{ThomasCover}, 
\begin{equation} \label{eq:MI2}
I(X,Y)\! =\! \sum_{x\in \mathcal{X},y\in \mathcal{Y}} p(x,y)\log_2 \left[ \frac{p(y\vert x)}{p(y)} \right].	
\end{equation}
This becomes evident when switching from the index notation to the notation in alphabet elements, that is, by replacing the transition probability matrix elements $A^j_i$ with the conditional probability $p(y\vert x)$, the joint probability $p^{ij}$ with $p(x,y)$, and the output probability $p^j$ with $p(y)$. The variable $x$ is selected from alphabet $\mathcal{X}$, and the variable $y$ is selected from the alphabet $\mathcal{Y}$ respectively.

Transfer entropy \cite{Schreiber} is an information theoretical implementation of ``Wieners principle of causality'' \cite{Wiener}: a ``cause'' combined with the past of the ``effect'' predicts the effect better than that the ``effect'' predicts itself. It was proven that with a slight modification of the original proposed transfer entropy (TE), it fully complies with Wieners principle of causality \cite{Wibral}. Transfer entropy is the measure of association for data transmission via a network of communication channels with an \emph{inverse multiplexer} topology (see Figure \ref{Multiplexer_djia}a). An inverse multiplexer comprises a demultiplexer and a multiplexer in series. The demultiplexer splits input data and transmits the data via one of several output lines. The multiplexer recombines the data from all output lines into one output line \cite{MUX}. This becomes apparent when transfer entropy is written as

\begin{equation*} \label{eq:TE_Multi}
 TE_{X\rightarrow Y} = \sum_{\psi^-_g\in \mathcal{Y}^{\ell}} p(\psi^-_g) \sum_{\substack{\textbf{x}^-\in \mathcal{X}^m\\
                  y\in \mathcal{Y}}\\} p(\textbf{x}^-,y|\textbf{y}^-\!=\! \psi^-_g) \log_2 \left[ \frac{p(y|\textbf{x}^-,\textbf{y}^-\!=\! \psi^-_g))}{p(y|\textbf{y}^-\!=\! \psi^-_g)} \right],
\end{equation*}

with $\textbf{x}^-$ representing the cause selecting symbols from the alphabet $\mathcal{X}^m$, $\textbf{y}^-$ representing the past of the effect, $\psi^-_g$ equals the $g^{th}$ alphabet element from the alphabet $\mathcal{Y}^{\ell}$, and $y$ representing the effect. It is assumed that $Y$ is a Markov process of order $\ell$, and that the ``cause'' consists of $m$ subsequent symbols. There are $\vert \mathcal{Y}^{\ell} \vert$ sub-channels, where $\vert \cdot \vert$ indicates the cardinality of the alphabet. Per sub-channel, the communication channel is identical to a communication channel giving rise to MI as the measure of association \cite{DSIG}, therefore, the elements of the $g^{th}$ sub-channel of the causal tensor $\mathcal{A}$ are given by ${A}^{{j}}_{g{i}}$. Like MI, transfer entropy can be expressed in terms of tensors:

\begin{equation*} \label{eq:TE_Tensor}
 TE_{X\rightarrow Y} = \sum_{g,{i},j} p^{{i}jg} \log_2 \left[ \frac{A^{j}_{g{i}}}{p^{j}_{g}} \right].
\end{equation*}

Mutual information also results from transmission over an inverse multiplexer, an inverse multiplexer comprising one sub-channel.

\subsubsection{Transmission Paths and Causal Tensors of a Cascade}
As mentioned in the introduction, a system comprising more than one random variable is represented by a graph. The vertices represent random variables, and the edges represent associations. The edges are undirected when MI is used as a measure of association and directed when TE is used. This gives rise to \emph{transmission paths}.
\begin{Definition}
	A transmission path, or path in short, is defined as the sequence in which edges have been used to transmit the data. The source is defined as the first variable in the path, the destination is the last variable in the path. A path is denoted as ``$\{ source\}\!\{mediator_1\}\! \cdots\!\{mediatior_n\}\!\{destination\}$''.
\end{Definition}
The association between a source and destination of a path comprising over two vertices, i.e., a cascade of (direct) paths, cannot be expressed in terms of the association constituting direct paths when MI or TE are used. In contrast, the resulting causal tensor can be expressed in terms of the constituting causal tensors \cite{DSIG}. For example, assume that the tensor elements for the path $\{x\}\!\{y\}$ are given by $A^{{j}}_{g{i}}$, with $g$ the index for the past of the effect ($y$), and the tensor elements for the path $\{y\}\!\{z\}$ are given by $B^{k}_{h{j}}$, with $h$ the index for the past of the effect ($z$). Where $\mathcal{T}\{x\}\!\{y\}\!\{z\}$ the causal tensor $\mathcal{T}$ for the path $\{x\}\!\{y\}\!\{z\}$ is equal to

\begin{equation} \label{eq:ChainTransmission}
	\mathcal{T}\{x\}\!\{y\}\!\{z\} = \sum_{g,j}p^g_{h{i}} A^{{j}}_{g{i}} B^{k}_{h{j}}.
\end{equation}

\begin{figure}[H]
\centering
\includegraphics[width=15 cm]{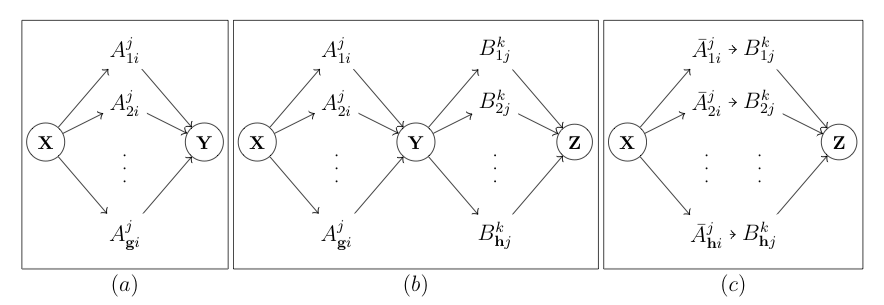} 
 \caption{\label{Multiplexer_djia} (\textbf{a}) The inverse multiplexer representing the communication network between $X$ and $Y$. Source data is partitioned on the past of the effect, indicated by index $g$, and transmitted via the related communication channel. (\textbf{b}) The inverse multiplexers representing the transmission the path $\{x\}\!\{y\}\!\{z\}$. Index $h$ is related to the past of the $z$. (\textbf{c}) An equivalent representation network communication channels representing the transmission the path $\{x\}\!\{y\}\!\{z\}$. The causal tensor element $\bar{A}^{{j}}_{h{i}}$ equals $\sum_{g}p^g_{h{i}} A^{{j}}_{g{i}}$ \cite{DSIG}. Per ``sub-channel'', the resulting causal tensor equals $\sum_j \bar{A}^{{j}}_{h{i}} {B}^{{k}}_{hj}$.}
\end{figure}

The term $\sum_{g}p^g_{h{i}} A^{{j}}_{g{i}}$ can be interpreted as the weighted sum of the causal tensors of the sub-channels of the first direct path, the path $\{x\}\!\{y\}$, and it evaluates to a causal tensor $\bar{A}^{{j}}_{h{i}}$ \cite{DSIG}. Figure \ref{Multiplexer_djia}b depicts the transmission of data over two inverse multiplexers in series, resulting in Equation (\ref{eq:ChainTransmission}). 
Transmission of data over a cascade of multichannel inverse multiplexers is equivalent to the transmission of data over a parallel set of cascades of mono-channel inverse multiplexers, i.e., results applicable to MI also apply to TE. For this reason, we can restrict ourselves, without loss of generality, to MI. Equation (\ref{eq:ChainTransmission}) then reduces to

\begin{equation} \label{eq:ChainTransmissionMI}
	\mathcal{T}\{x\}\!\{y\}\!\{z\} =\sum_{j} A^j_i B^k_j.
\end{equation}

Using this equation, the mutual information for the path $\{x\}\!\{y\}\!\{z\}$, $I\{x\}\!\{y\}\!\{z\}$, equals
\begin{equation*}
	I\{x\}\!\{y\}\!\{z\} = \sum_{i,j,k} p^{ijk}\log_2 \left[ \frac{\sum_j A^{j}_{i}B^{k}_{j}}{p^k} \right]. 
\end{equation*}

In the rest of this article we, sometimes, indicate the tensors of specific paths by there ``name'': $\mathcal{A}\!:=\!\mathcal{T}\{x\}\{y\}$, $\mathcal{A}^{\ddagger}\!:=\!\mathcal{T}\{y\}\{x\}$, $\mathcal{B}\!:=\!\mathcal{T}\{y\}\{z\}$, $\mathcal{B}^{\ddagger}\!:=\!\mathcal{T}\{z\}\{y\}$, $\mathcal{C}\!:=\!\mathcal{T}\{x\}\{z\}$, and $\mathcal{C}^{\ddagger}\!:=\!\mathcal{T}\{z\}\{x\}$. Using these ``names'', it is proved in Appendix \ref{App:TraversePath} that the mutual information of a path is symmetric.

\begin{Lemma} \label{col:opposite_traverse}
	The mutual information of a path equals the mutual information of the path traversed in the opposite direction:
	\begin{equation}
		I\{ x\}\{y\}\{z\} = I\{z\}\{y\}\{ x\}.
	\end{equation}
\end{Lemma}

\subsubsection{Indirect Associations and No Associations}
There are two underlying reasons to delete an edge in a graph: \textit{(i)} the association it represents is indirect, or \textit{(ii)} there is no association at all. These two reasons can be distinguished using causal tensors. First, because the causal tensor of a cascade can be determined exactly from the constituting causal tensors, causal tensors can differentiate between direct and indirect associations \cite{DSIG}. Second, if there is no association between two vertices at all, the causal tensor represents a communication channel that cannot transmit any information. In this case the transition probability matrix has identical rows, e.g., $\forall\;i\neq f:\; A^j_i = A^j_f$. 

\begin{Proposition}
	If the association between two vertices is indirect, or if the association does not exist, the direct path does not exist. For example, the path $\{x\}\!\{z\}$ does not exist when the graph $X\!\rightarrow\! Y\!\rightarrow\!Z$ is the ground truth. 
\end{Proposition}

%
\subsection{Partial Information Decomposition} \label{ss:PID}
The partial information decomposition framework of Williams and Beer \cite{PID} allows for a decomposition of the total information in non-negative unique, redundant, and synergistic information components. The unique information $\mathcal{U}(Y;Z)$ represents information in $Z$ only provided by $Y$ and not by $X$. The redundant information $\mathcal{R}(X,Y;Z)$ represents the information in $Z$ provided by both $X$ and $Y$. The synergistic information $\mathcal{S}(X,Y;Z)$ represents information in $Z$ that results via interaction between $X$ and $Y$. The relations between these information components for a system comprising three variables are given by the following set of equations,
 
\begin{align} 
	I(X,Y;Z) &= \mathcal{U}(Y;Z)+\mathcal{U}(X;Z)+\mathcal{R}(X,Y;Z)+\mathcal{S}(X,Y;Z), \label{PID1}\\
	I(Y;Z) &= \mathcal{U}(Y;Z)+\mathcal{R}(X,Y;Z), \label{PID2}\\
	I(X;Z) &= \mathcal{U}(X;Z)+\mathcal{R}(X,Y;Z). \label{PID3}
\end{align}

Williams and Beer propose three redundancy related axioms: \textit{(i)} Symmetry: redundancy does not change when sources are permuted, e.g., $\mathcal{R}(X,Y;Z)\! = \! \mathcal{R}(Y,X;Z)$. \textit{(ii)} Self-Redundancy: for a single source, the redundancy equals the mutual information between the source and the target, e.g., $\mathcal{R}(X,X;Z)\! = \! I(X;Z)$. From now on we set $\mathcal{R}(X,X;Z)\!=\! \mathcal{R}(X;Z)$ \textit{(iii)} Monotonicity: the redundancy does not increase when a new source is added, e.g., $\mathcal{R}(X;Z)\! \geq \! \mathcal{R}(X,Y;Z)$.

\subsubsection{Redundancy, Indirect Paths and the Data Processing Inequality}
If the association between two vertices is indirect, no direct path exists. Source data is transmitted to the mediator vertex, which  stores, possibly copies, possibly modifies, and possibly enriches the received information with information from this mediator vertex, after which it is transmitted again towards the next mediator vertex or destination, i.e., the target. This consideration leads to the following proposition:

\begin{Proposition}
Unique information can only result from data transmission via a direct path. Redundant information is the consequence of data transmission via an indirect path.
\end{Proposition}

A direct consequence of this proposition is that in case the chain $X\!\rightarrow\! Y\!\rightarrow\!Z$ is the ground truth, no unique information is shared between $X$ and $Z$, or stated otherwise, all information shared between $X$ and $Z$ is redundant. Another immediate consequence of this proposition is that in case of the \textsc{exclusive or} example from \cite{XOR}, there is neither unique, nor redundant information shared between the sources and the target: all information shared is synergistic. 

In a fully connected three-node system, there are two source vertices transmitting data to the target vertex, and, per source vertex, there is one indirect path between that source vertex and the target vertex. The Data Processing Inequality (DPI) \cite{ThomasCover} establishes an inequality between the indirect path and the constituting direct paths. Data Processing Inequality states that processing of data can never increase the amount of information. The DPI applied to the path $\{x\}\!\{y\}\!\{z\}$ results in $I\{x\}\!\{y\}\!\{z\}\! \leq\! \min [I\{x\}\!\{y\}\!,I\{y\}\!\{z\}]$. Via this path, the redundant information from $X$ is transmitted. Applied to the path $\{y\}\!\{x\}\!\{z\}$, the DPI results in $I\{y\}\!\{x\}\!\{z\}\! \leq\! \min [I\{y\}\!\{x\}\!,I\{x\}\!\{z\}]$. Based on this example, the following definition of path based redundancy is proposed:

\begin{Definition} \label{def:Redundancy} Path based redundant information shared between a specific set of sources with respect to a target is defined as the weakest of all indirect paths that: (1) contain all the sources, (2) start with a source and ends at the target, and (3) do not contain non-existing paths. For a system comprising three variables,

\begin{equation} \label{eq:Redundancy3}
	\mathcal{R}(X,Y;Z) = \min \left[ I\{x\}\{y\}\{z\},I\{y\}\{x\}\{z\}  \right].
\end{equation}
	
\end{Definition} 
Because the redundancy in this definition equals a mutual information, the proposed path based redundancy is per definition non-negative. 
\subsection{Characteristics of Path Based Redundancy}
Using Definition \ref{def:Redundancy} and Lemma \ref{col:opposite_traverse}, it is proven in Appendix \ref{App:SymmetryOfRedundancy} that

\begin{Theorem} \label{Thm:SymmetryOfRedundancy}

\begin{equation}
	\mathcal{R}(X,Y;Z) \geq \mathcal{R}(Z,Y;X) \textsc{ or } \mathcal{R}(X,Y;Z) \geq \mathcal{R}(Z,X;Y).
\end{equation}
	
\end{Theorem}

\subsubsection{Symmetry, Self-Redundancy, and Monotonicity}
The path based redundancy from Definition \ref{def:Redundancy} also satisfies the three axioms introduced in \cite{PID}.
\begin{proof}[Proof of Symmetry]
The symmetry $\mathcal{R}(X,Y;Z)\! = \! \mathcal{R}(Y,X;Z)$ follows directly from Definition \ref{def:Redundancy}: all indirect paths representing all permutations of sources are compared. The order of the sources in the redundancy expression is therefore irrelevant.
\end{proof}
\begin{proof}[Proof of Self-Redundancy]
Consider a system consisting of three variables, the sources $X$ and $Y$, and the target $Z$. Let now assume that one source is a copy of the other source, i.e., $Y\!=\!X$. This means that the causal tensors describing the mapping between the sources and target, $\mathcal{T}\{y\}\{z\}$ and $\mathcal{T}\{x\}\{z\}$, are equal: $\mathcal{T}\{y\}\{z\}=\mathcal{T}\{x\}\{z\}$. Per definition $\mathcal{R}(X;Z)\! =\! I\{x\}\!\{x\}\!\{z\}$. The causal tensor for the path $\{x\}\!\{x\}$, $\mathcal{T}\{x\}\{x\}$, equals the Kronecker delta $\delta^j_i$: $\delta^j_i \!=\! 0$ unless $i\!=\! j$ in which case $\delta^j_i\! =\! 1$. Therefore, using Equation (\ref{eq:ChainTransmission}), the causal tensor for the path $\{x\}\!\{x\}\!\{z\}$ equals the tensor $\mathcal{C}$ for the path $\{x\}\!\{z\}$,  i.e., $I\{x\}\!\{x\}\!\{z\} = I(X;Z)$, in other words, for a single source, the redundancy equals the mutual information between the source and the target. 
\end{proof}
\begin{proof}[Proof of Monotonicity]
Because of the definition of redundancy, adding more sources never decreases the number of edges in the cascade: $\mathcal{R}(X;Z)\! \geq \! \mathcal{R}(X,Y;Z)$. As per Data Processing Inequality, adding sources can therefor never increase the redundancy.
\end{proof}
\subsubsection{The Redundancy Lattice, Transmission Paths and the DPI}
\begin{figure}[H]
\centering
\includegraphics[width=12 cm]{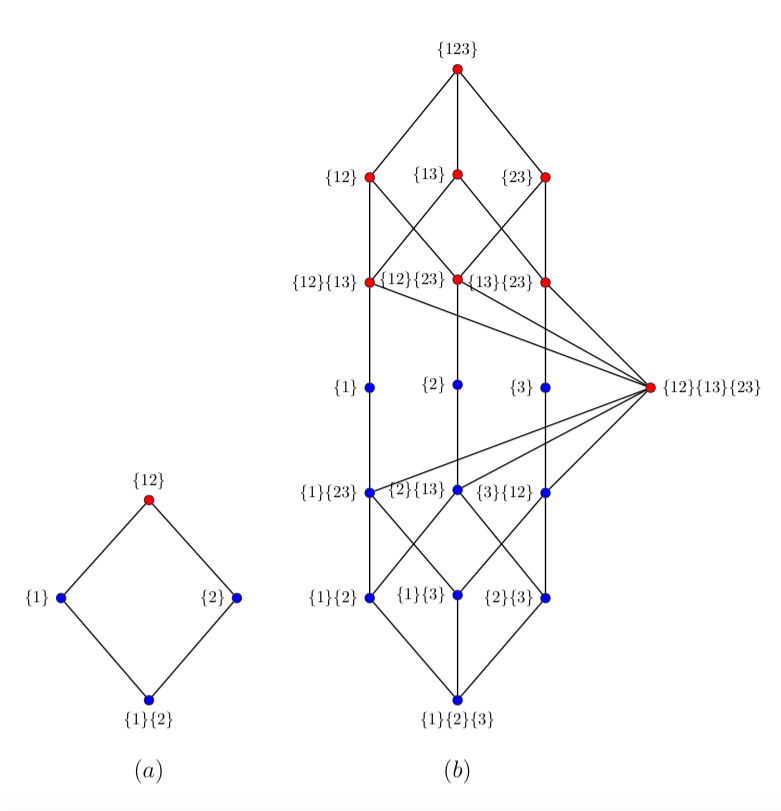} 
 \caption{\label{RLattice} The redundancy lattice \cite{PID} for: (\textbf{a}) two sources, and (\textbf{b}) three sources. In case two lattice vertices are connected, the redundancy related to the highest lattice vertex (in position) is greater than or equal to the redundancy of the lower lattice vertex. The ordering relation for the vertices of the same color follows directly from the Data Processing Inequality.}
\end{figure}

In Figure \ref{RLattice} the redundancy lattices from \cite{PID} are depicted. These lattices reflect the ordering relation in the redundancy.
The target vertex is omitted in the lattice labels. If we would add the target vertex, say $\{t\}$, the labels represent all the transmission paths resulting from permutation of the sources. The \emph{DPI invokes an ordering relation}. For example, the vertex with the label $\{ 1 \} \{ 2 \}$ in Figure \ref{RLattice}a, represents the redundancy related to the paths $\{ 1 \} \{ 2 \} \{ t \}$ and $\{ 2 \} \{ 1 \} \{ t \}$. The vertex with the label $\{ 1 \}$  represents the redundancy related to the path $\{ 1 \}\{ t \}$. Let now assume that, without loss of generality, $\mathcal{R}(1,2;t) = I\{ 1 \} \{ 2 \}\{ t \}$.  The Data Processing Inequality gives us $I\{ 1 \} \{ 2 \}\{ t \} \leq \min \left[ I\{ 1 \}\{ 2 \}, I\{ 2 \}\{ t \} \right]$, that is, $\mathcal{R}(1,2;t)\! \leq\! \mathcal{R}(1;t)$, and $\mathcal{R}(1,2;t)\! \leq\! \mathcal{R}(2;t)$. In \cite{DSIG} is is proven that the tensor of the path $\{ 1\}\{t \}$ is equal to the tensor of the path $\{ 1\} \{12 \}\{t\}$, and that the tensor of the path $\{ 2\}\{t \}$ equals the tensor of the path $\{ 2\} \{12 \}\{t\}$. The DPI now results in the ordering relations $\mathcal{R}(1;t)\! \leq\! \mathcal{R}(12;t)$, and $\mathcal{R}(2;t)\! \leq\! \mathcal{R}(12;t)$. To summarize, for a system comprising three variables, the proposed redundancy definition results in the ordering relations that matches the redundancy lattice derived in \cite{PID}:  
\begin{itemize}
\setlength\itemsep{0em}
	\item $\mathcal{R}(1,2;t)\! \leq\! \mathcal{R}(1;t)$,
	\item $\mathcal{R}(1,2;t)\! \leq\! \mathcal{R}(2;t)$,
	\item $\mathcal{R}(1;t)\! \leq\! \mathcal{R}(12;t)$,
	\item $\mathcal{R}(2;t)\! \leq\! \mathcal{R}(12;t)$.
\end{itemize}

For a system comprising three sources, the DPI also results in an ordering relation. To facilitate the application of the DPI, the vertices have been colored blue or red (see Figure \ref{RLattice}b). For the blue vertices (the bottom four levels)the redundancy associated with a lattice vertex must be less than or equal to the redundancy of lattice vertices comprising only a subset of the sources as per DPI. For example, $\mathcal{R}(1,23;t)\! \leq\! \mathcal{R}(1;t)$. When this is combined with the order suggested by the redundancy lattice for two sources, e.g., $\mathcal{R}(12;t)\!\geq\! \mathcal{R}(1;t)$, the proposed definition fully complies with the order implied by the blue vertices. The same is true for the red vertices (the top three levels, including the vertex with the label $\{12\}\{13\}\{23\}$. The ordering relation between connected blue and red vertices, e.g., $\{ 1\}$ and $\{12\}\{13\}$ in Figure \ref{RLattice}a, follows, as earlier mentioned, from \cite{DSIG}.
\subsubsection{Identity and Left Monotonicity}
Apart from these three axioms, Bertschinger et al. \cite{LeftChainRule} and Harder et al. \cite{Identity} proposed other properties. The proposed redundancy measure satisfies the ``identity property''. The intuition behind the identity property \cite{Identity} is that if the target is a join of the inputs, the redundancy equals the mutual information of inputs, i.e.,
\begin{equation} \label{eq:Identity}
 \mathcal{R}(X,Y;XY) = I(Y;X), 
\end{equation}
with $XY \!=\! X\cup Y$. In Appendix \ref{App:Identity} we prove that the proposed redundancy property satisfies the identity property. 

The Left monotonicity property captures the intuition that the redundancy does not \emph{decrease} when a new destination is added \cite{LeftChainRule}. In Appendix \ref{App:LeftMono} it is proven that the proposed redundancy satisfies the left monotonicity axiom
\begin{equation} \label{eq:LeftMono}
	\mathcal{R}(X,Y;Z_1Z_2) \geq \mathcal{R}(X,Y;Z_1).
\end{equation}
\subsubsection{Unique Information}
Unique information for a three-node system is defined by Equation (\ref{PID2}) and Equation (\ref{PID3}). Using the proposed redundancy. Using Definition \ref{def:Redundancy}it follows immediately that the unique information is non-negative. The resulting unique information fully matches our intuition. This is demonstrated by with the fully connected three-node system. Redundant information in $Z$ results from information transmitted via the paths $\{x\}\!\{y\}\!\{z\}$ and $\{y\}\!\{x\}\!\{z\}$. The causal tensor elements of the path $\{y\}\!\{x\}\!\{z\}$ and the path $\{x\}\!\{y\}\!\{z\}$ are given by $\sum_{i} A^{\ddagger i}_j C^k_i$ and $\sum_{j} A^{j}_i B^k_j$ respectively. Using Equation (\ref{PID2}), the unique information equals
\begin{equation}\label{YUnique}
	\mathcal{U}(Y;Z) = \max \Biggl[ \sum_{i,j,k} p^{ijk} \log_2 \Biggl[ \frac{B^{k}_j}{\sum_m A^{\ddagger m}_{j}C^{k}_{m}} \Biggl], \; \sum_{i,j,k} p^{ijk} \log_2 \Biggl[ \frac{B^{k}_j}{\sum_\ell A^{\ell}_{i}B^{k}_{\ell}} \Biggl]\Biggl].
\end{equation} 
The first sum is a measure for the divergence between the direct the path $\{y\}\!\{z\}$ and the indirect path $\{y\}\!\{x\}\!\{z\}$. If the \emph{association} between $Y$ and $Z$ is indirect, this sum evaluates to zero, i.e., the direct path between $Y$ and $Z$ does not exist. In this case the tensor $\mathcal{B}$ represents a communication channel that can not transmit any information: the second sum also equals zero. The second sum is an indication of how much $Y$ differs from $X$. If $Y$ is an exact copy of $X$ the tensor $\mathcal{A}$ equals the Kronecker delta, this second sum evaluates to zero. In that case, $\mathcal{A}^{\ddagger} \!=\! \mathcal{C}$ and $\mathcal{B}\!=\! \mathcal{C}$, i.e., the first sum also vanishes.
\section{Results} \label{s:Results}
In this section we start with investigating the behavior of the proposed partial information decomposition with respect the conceptual issue related to, $I_{min}$, the original redundancy measure used in \cite{PID}: $I_{min}$ does not distinguish between ``same information'' or ``the same amount of information'', i.e., it conflicts the identity property \cite{Identity}. We already know that the proposed redundancy satisfies the identity relation, but this example illustrates the proposed method and it familiarizes the reader with the method.
\subsection{Two Bit Copy Problem} \label{ss:TBC}
A conceptual problem with the redundancy measure used in \cite{PID}, $I_{min}$, is illustrated with the so called ``two-bit copy problem''. For two independent and identically distributed binary variables $X$ and $Y$, the  target $Z$ is a copy of these two variables: $Z = (X, Y)$. It can be shown that $I_{min}\{1\}\{2\}\{Z\}\!=1$ bit \cite{Ince}. The problem lies in the fact that there is no overlap between the information of both variables: the result does not match out intuition, $I_{min}$ seems to overestimate the redundancy.
{ 
\renewcommand{\arraystretch}{1.3}
\begin{table}[h]
\caption{Distribution for the ``two-bit copy problem''.}
\centering
\begin{tabular}{cccc}
\toprule
\emph{X} & \emph{Y} & \emph{Z} 	& \emph{p(Z)}\\
\hline
0 & 0 & (0,0)  & $\tfrac{1}{4}$ \\
0 & 1 & (0,1)  & $\tfrac{1}{4}$ \\
1 & 0 & (1,0)  & $\tfrac{1}{4}$ \\
1 & 1 & (1,1)  & $\tfrac{1}{4}$ \\
\hline
\end{tabular}
\label{tbl:TruthCopy}
\end{table}
}
The proposed PID does not suffer from this issue. This follows immediately from our definitions and the fact that there are no paths between $X$ and $Y$ because the variables are independent. However, for illustrative purpose we use a general approach. First the graph representing this system has to be inferred, and the indirect associations have to be removed. For this bivariate causal tensors are used \cite{DSIG}. The tensors and the probability mass functions (pmfs) are then used to determine the redundancy. We start with determining the tensor for the path $\{x\}\!\{y\}\!\{z\}$. This path is a cascade of two paths, $\{x\}\{y\}$ and $\{y\}\{z\}$. The resulting tensor is the (matrix) product of the two related tensors $\mathcal{T}\{x\}\{y\}$ and $\mathcal{T}\{y\}\{z\}$ respectively. From Table \ref{tbl:TruthCopy} it follows that

\small{
\begin{equation*}
\mathcal{T}\{x\}\{y\} = 
\begin{pmatrix} 
\tfrac{1}{2} & \tfrac{1}{2}\\
\tfrac{1}{2} & \tfrac{1}{2}  
\end{pmatrix},\text{ and }
\mathcal{T}\{y\}\{z\} = 
\begin{pmatrix} 
\tfrac{1}{2} & 0 & \tfrac{1}{2} & 0\\
0 & \tfrac{1}{2} & 0 & \tfrac{1}{2}  
\end{pmatrix}.
\end{equation*}
}\normalsize

As per Equation (\ref{eq:ChainTransmissionMI}), the tensor for the path $\{x\}\!\{y\}\!\{z\}$ equals
\small{
\begin{equation*}
\mathcal{T}\{x\}\!\{y\}\!\{z\} = 
\begin{pmatrix} 
\tfrac{1}{4} & \tfrac{1}{4} & \tfrac{1}{4} & \tfrac{1}{4}\\
\tfrac{1}{4} & \tfrac{1}{4} & \tfrac{1}{4} & \tfrac{1}{4}  
\end{pmatrix}.
\end{equation*}
}\normalsize

This tensor contains the transition probabilities, i.e., the conditional probabilities $p(z \vert x)$. Applying this tensor to the probability mass function for the input $X$, $p(x)\! =\! (\tfrac{1}{2},\tfrac{1}{2})$, results in
 
\begin{equation*}
	p(x) \cdot \mathcal{T}\{x\}\!\{y\}\!\{z\} = (\tfrac{1}{4}, \tfrac{1}{4}, \tfrac{1}{4}, \tfrac{1}{4}),
\end{equation*}

the pmf for the output $Z$. For this path $I\{x\}\!\{y\}\!\{z\} = 0$. We used Equation (\ref{eq:MI}) and the pmfs and tensor $\mathcal{T}\{x\}\!\{y\}\!\{z\}$ as compiled in Table \ref{tbl:TruthCopyPMF}. 
{ 
\renewcommand{\arraystretch}{1.3}
\begin{table}[H]
\caption{Marginal, conditional and joint distributions for path $\{x\}\!\{y\}\!\{z\}$, with $p(Z\vert X)=\mathcal{T}\{x\}\!\{y\}\!\{z\}$, and $p(Z) = p(X) \cdot \mathcal{T}\{x\}\!\{y\}\!\{z\}$.}
\centering
\begin{tabular}{ccccccc}
\toprule
\emph{Source X}  & \emph{Target Z} 	& \emph{p(X)}  & \emph{p(Z)} & \emph{$p(Z\vert X)$} & \emph{$p(Z,X)$} & \emph{$p(Z\vert X)$}\Big/\emph{p(Z)}\\
\hline
0 &  (0,0)  & $\tfrac{1}{2}$ & $\tfrac{1}{4}$ & $\tfrac{1}{4}$ & $\tfrac{1}{8}$ & 1\\
0 &  (0,1)  & $\tfrac{1}{2}$ & $\tfrac{1}{4}$ & $\tfrac{1}{4}$ & $\tfrac{1}{8}$ & 1\\
0 &  (1,0)  & $\tfrac{1}{2}$ & $\tfrac{1}{4}$ & $\tfrac{1}{4}$ & $\tfrac{1}{8}$ & 1\\
0 &  (1,1)  & $\tfrac{1}{2}$ & $\tfrac{1}{4}$ & $\tfrac{1}{4}$ & $\tfrac{1}{8}$ & 1\\
1 &  (0,0)  & $\tfrac{1}{2}$ & $\tfrac{1}{4}$ & $\tfrac{1}{4}$ & $\tfrac{1}{8}$ & 1\\
1 &  (0,1)  & $\tfrac{1}{2}$ & $\tfrac{1}{4}$ & $\tfrac{1}{4}$ & $\tfrac{1}{8}$ & 1\\
1 &  (1,0)  & $\tfrac{1}{2}$ & $\tfrac{1}{4}$ & $\tfrac{1}{4}$ & $\tfrac{1}{8}$ & 1\\
1 &  (1,1)  & $\tfrac{1}{2}$ & $\tfrac{1}{4}$ & $\tfrac{1}{4}$ & $\tfrac{1}{8}$ & 1\\
\hline
\end{tabular}
\label{tbl:TruthCopyPMF}
\end{table}
}
Because $\mathcal{R}(X,Y;Z)\!=\! \min \left[I\{x\}\!\{y\}\!\{z\}, I\{y\}\!\{x\}\!\{z\} \right]$ we find that there is no redundant information contribution.
\subsection{Negative contributions}
Rauh et al. \cite{LocalPositivity} demonstrated that left monotonicity property and the identity property are incompatible with nonnegative partial information components for systems comprising over three nodes, i.e., the \emph{local positivity} property is violated. In this section it is demonstrated that for the proposed PID, negative contributions are also possible for systems comprising three variables. We furthermore offer an explanation as to why negative contributions can arise in these cases. 
\subsubsection{Negative Synergistic Contributions Due to Unobserved Common Causes} \label{Section:Really}
A result of the proposed definitions is that an unobserved common cause can lead to a negative synergistic contribution. In Table \ref{tbl:NegativeSynergy} an example is given of such a system. The related graphs are depicted in Figure \ref{Unobserved}. In Appendix \ref{App:UnObserved} it is shown that in there is no path between $X$ and $Y$ in case $\tilde{Z}$ is observed.

{ 
\renewcommand{\arraystretch}{1.3}
\begin{table}[H]
\caption{Example of a system with an unobserved common cause. \textbf{(a)} Data set comprising three parameters. \textbf{(b)} Hidden common cause. $X=\tilde{Z}_1$, $Y= \tilde{Z}_1 \textsc{Or} \tilde{Z}_2$, and $Z = \tilde{Z}_1 \textsc{And} \tilde{Z}_2$.}
\centering
\begin{tabular}{cccccc}
\toprule
\textbf{} & \textbf{(a) Data set} & \textbf{} 	&  \textbf{} & \textbf{(b) Common cause} & \textbf{} \\
\hline
\emph{X} & \emph{Y} & \emph{Z} 	& \textbf{} & $\tilde{Z}=\tilde{Z}_1\tilde{Z}_2$ &  \emph{$p(\tilde{Z})$}  \\
\hline
0 & 0 & 0  & & 00 & $\tfrac{1}{4}$ \\
0 & 1 & 0  & & 01 & $\tfrac{1}{4}$ \\
1 & 1 & 0  & & 10 & $\tfrac{1}{4}$ \\
1 & 1 & 1  & & 11 & $\tfrac{1}{4}$ \\
\hline
\end{tabular}
\label{tbl:NegativeSynergy}
\end{table}
} 
If the common cause $\tilde{Z}$ is observed, the path $\{ x\}\{y \}$ does not exist when the tensor of this path equals the tensor of the path $\{ x\}\{\tilde{z} \}\{y \}$. 
\begin{figure}[H]
\centering
\includegraphics[width=7 cm]{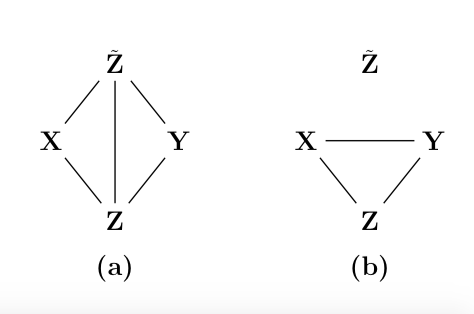} 
 \caption{\label{Unobserved} \textbf{(a)} The graph when all variables are observed. \textbf{(b)} The graph when $\tilde{Z}$ is not observed.}
\end{figure}

Using Table \ref{tbl:NegativePMFXZ} and Equation (\ref{eq:MI}), we find that $\mathcal{R}(X,Y;Z)\! \leq\! 0.0271bit$. Because the interaction information, $I(X;Y\vert Z)\! -\! I(X;Y)$ \cite{InteractionInformation}, equals $-0.1226bit$, and  $I(X;Y\vert Z)\! -\! I(X;Y)\! =\! \mathcal{S}(X,Y;Z)-\mathcal{R}(X,Y;Z)$ \cite{PID}, the synergistic information is negative.\\ 

The proposed PID can be used to infer the existence of a non-observed variable. For a system comprising three variables this will always be the case when

\begin{equation}
	I(X;Y\vert Z) - I(X;Y) \leq -\min \left[I\{x\}\!\{y\}\!\{z\}, I\{y\}\!\{x\}\!\{z\} \right].
\end{equation} 
{ 
\renewcommand{\arraystretch}{1.3}
\begin{table}[H]
\caption{Marginal, conditional and joint distributions for path $\{x\}\!\{y\}\!\{z\}$, with $p(Z\vert X)=\mathcal{T}\{x\}\!\{y\}\!\{z\}$, and $p(Z) = p(X) \cdot \mathcal{T}\{x\}\!\{y\}\!\{z\}$.}
\centering
\begin{tabular}{ccccccc}
\toprule
\emph{Source X}  & \emph{Target Z} 	& \emph{p(X)}  & \emph{p(Z)} & \emph{$p(Z\vert X)$} & \emph{$p(Z,X)$} & \emph{$p(Z\vert X)$}\Big/\emph{p(Z)}\\
\hline
0 &  0  & $\tfrac{1}{2}$ & $\tfrac{3}{4}$ & $\tfrac{5}{6}$ & $\tfrac{5}{12}$ & $\tfrac{10}{9}$\\
0 &  1  & $\tfrac{1}{2}$ & $\tfrac{1}{4}$ & $\tfrac{1}{6}$ & $\tfrac{1}{12}$ & $\tfrac{2}{3}$\\
1 &  0  & $\tfrac{1}{2}$ & $\tfrac{3}{4}$ & $\tfrac{2}{3}$ & $\tfrac{4}{12}$ & $\tfrac{8}{9}$\\
1 &  1  & $\tfrac{1}{2}$ & $\tfrac{1}{4}$ & $\tfrac{1}{3}$ & $\tfrac{2}{12}$ & $\tfrac{4}{3}$\\
\hline
\end{tabular}
\label{tbl:NegativePMFXZ}
\end{table}
} 
\subsubsection{Negative Contributions Due to Overestimation of Unique Information}
\begin{figure}[H]
\centering
\includegraphics[width=8 cm]{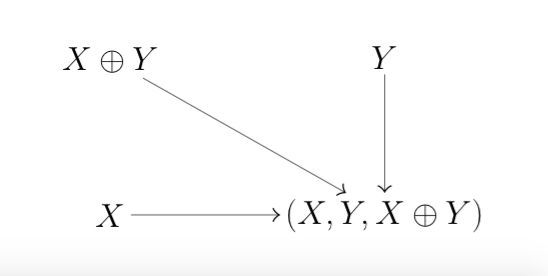} 
 \caption{\label{LocalPositivity} The graph for the system used to demonstrate that the left monotonicity property and the identity property are incompatible with the local positivity property.}
\end{figure}

The system used by Rauh et al. \cite{LocalPositivity} comprises three source nodes: $X$, $Y$, and $X \oplus Y$. The sources $X$ and $Y$ are independent identically distributed binary variables. The target node is defined as $(X, Y, X \oplus Y)$. In Figure \ref{LocalPositivity} the underlying graph is depicted. This structure follows directly from the independence of $X$ and $Y$, and from the probability distribution of the \textsc{Xor}. While in the previous example negative arose due to an \emph{underdetermined system}, i.e., a system with unobserved variables, we hypothesize that negativity in this case is due to an \emph{overdetermined} system.

The system in Figure \ref{LocalPositivity} only contains unique and synergistic contributions. Irrespective of the contributions of the lower three levels in Figure \ref{RLattice}b, the unique contributions indicated by the nodes $\{1\}$, $\{2\}$ and $\{3\}$ add up to \emph{3 bit}, while the target can be described by \emph{2 bit}. this implies that there must be a negative contribution. In all cases where: \textit{(i)} an extra source is added that is independent from all other sources, and \textit{(ii)}, the same variable is added to the target, one could expect an overestimation of the unique information and therefore a potential negative contribution.\\

The two examples discussed in this section demonstrate that the proposed PID can result in negative information contributions. What these examples also demonstrate, is that negativity of the information components can be interpreted as violations of assumptions. For the system comprising three variables, a negative synergistic contribution arises due to the incorrect assumption that all sources are known. In the example of the system comprising four variables, a negative term can arise when the assumption about what constitutes a source, and a target is incorrect. If these examples describe all causes of negative contributions is an open question that warrants future research.
\subsection{Dyadic and Triadic Systems}
Next we apply the proposed method to the two data sets from Table \ref{tbl:BeyondShannon}. Although these sets have different underlying dependency structures, they apparently have the same statistical structure \cite{BeyondShannon}.
{ 
\renewcommand{\arraystretch}{1.3}  
\begin{table}[H]
\caption{Two systems, both comprising three random variables with identical joint probabilities per combination of the random variables. The underlying structures are very different, which can be seen when the variables are represented in two bits, e.g., the binary expansion for $X\!=\!3$ equals $X_0X_1\!=\! 11$. \textbf{(a)} For the dyadic (pair-wise) set, $X_0\!=\!Y_1$,$Y_0
\!=\!Z_1$, and $Z_0\!=\!X_1$. \textbf{(b)} For the triadic (three-way) set, $X_0+Y_0+Z_0$ mod2, and $X_1+Y_1+Z_1$.}
\centering
\begin{tabular}{cccccccccc}
\toprule
\textbf{} & \textbf{(a) Dyadic} & \textbf{} 	&  \textbf{} & \textbf{} & \textbf{} & \textbf{} & \textbf{(b) Triadic}	& \textbf{} &  \textbf{} \\
\hline
\emph{X} & \emph{Y} & \emph{Z} 	&  \emph{p} & \textbf{} & \textbf{} & \emph{X} & \emph{Y} & \emph{Z} &  \emph{p} \\
\hline
0 & 0 & 0  & $\tfrac{1}{8}$ & & & 0 & 0 & 0  & $\tfrac{1}{8}$\\
0 & 2 & 1  & $\tfrac{1}{8}$ & & & 1 & 1 & 1  & $\tfrac{1}{8}$\\
1 & 0 & 2  & $\tfrac{1}{8}$ & & & 0 & 2 & 2  & $\tfrac{1}{8}$\\
1 & 2 & 3  & $\tfrac{1}{8}$ & & & 1 & 3 & 3  & $\tfrac{1}{8}$\\
2 & 1 & 0  & $\tfrac{1}{8}$ & & & 2 & 0 & 2  & $\tfrac{1}{8}$\\
2 & 3 & 1  & $\tfrac{1}{8}$ & & & 3 & 1 & 3  & $\tfrac{1}{8}$\\
3 & 1 & 2  & $\tfrac{1}{8}$ & & & 2 & 2 & 0  & $\tfrac{1}{8}$\\
3 & 3 & 3  & $\tfrac{1}{8}$ & & & 3 & 3 & 1  & $\tfrac{1}{8}$\\
\hline
\end{tabular}
\label{tbl:BeyondShannon}
\end{table}
} 
For the dyadic set, the causal tensors are given by:
\footnotesize{
\begin{equation*}
\mathcal{T}\{x\}\{y\} = 
\begin{pmatrix} 
\tfrac{1}{2} & 0 & \tfrac{1}{2} &0   \\
\tfrac{1}{2} & 0 & \tfrac{1}{2} &0   \\
 0 & \tfrac{1}{2} & 0 & \tfrac{1}{2} \\
 0 & \tfrac{1}{2} & 0 & \tfrac{1}{2}
\end{pmatrix}
,\;
\mathcal{T}\{x\}\{z\} = 
\begin{pmatrix} 
\tfrac{1}{2} & \tfrac{1}{2} & 0 &0   \\
0 & 0 & \tfrac{1}{2} & \tfrac{1}{2}  \\
\tfrac{1}{2} & \tfrac{1}{2} & 0 & 0   \\
0 & 0 & \tfrac{1}{2} & \tfrac{1}{2}
\end{pmatrix},
\end{equation*}
}\normalsize
$\mathcal{T}\{y\}\{z\}\! =\! \mathcal{T}\{x\}\{y\}, \; \mathcal{T}\{y\}\{x\}\! =\! \mathcal{T}\{x\}\{z\},\; \mathcal{T}\{z\}\{y\}\! =\! \mathcal{T}\{x\}\{z\}$, and $\mathcal{T}\{z\}\{x\}\! =\! \mathcal{T}\{x\}\{y\}$. Because no relation is the result of a cascade, e.g., $\mathcal{T}\{y\}\{z\}\! \neq\! \mathcal{T}\{y\}\{x\} \cdot \mathcal{T}\{x\}\{z\}$, the structure is that of an undirected triangle. Lets assume we are interested in the partial information decomposition of the total information in $Z$. The reader can confirm that all the elements of the tensor $\mathcal{T}\{x\}\{y\}\{z\}$ equal $\tfrac{1}{4}$: there is no redundant information contribution to $Z$.\\

For the triadic set, the causal tensors are given by:
\footnotesize{
\begin{equation*}
\mathcal{T}\{x\}\{y\} = 
\begin{pmatrix} 
\tfrac{1}{2} & 0 & \tfrac{1}{2} &0   \\
0 & \tfrac{1}{2} & 0 & \tfrac{1}{2}  \\
\tfrac{1}{2} & 0 & \tfrac{1}{2} &0   \\
0 & \tfrac{1}{2} & 0 & \tfrac{1}{2} 
\end{pmatrix},
\end{equation*}
}\normalsize
$\mathcal{T}\{y\}\{z\}\! =\! \mathcal{T}\{x\}\{y\}, \; \mathcal{T}\{x\}\{z\}\! =\! \mathcal{T}\{x\}\{y\}, \;\mathcal{T}\{y\}\{x\}\! =\! \mathcal{T}\{x\}\{y\},\; \mathcal{T}\{z\}\{y\}\! =\! \mathcal{T}\{x\}\{y\}$, and $\mathcal{T}\{z\}\{x\}\! =\! \mathcal{T}\{x\}\{y\}$. Here, the ground structure is that of a chain because any relation results from a cascade, e.g., $\mathcal{T}\{x\}\{z\}\! =\! \mathcal{T}\{x\}\{y\} \cdot \mathcal{T}\{y\}\{z\}$. Let now for example investigate the chain $X\! \rightarrow\! Y\! \rightarrow\! Z$. By definition, $Z$ only contains redundant information from $X$.\\

This example shows that the difference in underlying structure is reflected in two ways. First, the graphs related to the dyadic set and the triadic set are different (a triangle versus a chain). Second, for the dyadic set there is no redundant information, while the triadic set, $Z$ does contain redundant information.
\subsection{Comparison with Other Measures} \label{ss:Compare}
To get an idea about the behavior of the proposed redundancy measure, we compared it with four other measures: \textit{(i)} Pointwise Partial Information Decomposition, using the average Partitial Information atoms $\langle i \rangle$, \textit{(ii)} the redundancy measure proposed by Williams and Beer $I_{min}$, \textit{(iii)} $I_{broja}$, the redundancy measure proposed in \cite{BRAJO}, and \textit{(iv)} the redundancy based on Pointwise Common Change in Surprisal, $I_{CCS}$\cite{Ince}. The proposed redundancy measure is represented as $I_\bigtriangleup$. 
\subsubsection{Comparison with Pointwise Partial Information Decomposition}
\begin{figure}[H]
\centering
\includegraphics[width=12 cm]{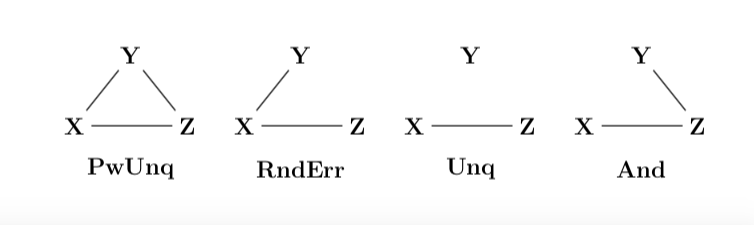} 
 \caption{\label{PPIDFour} The underlying structures for the probability distributions of Table \ref{tbl:PPID}.}
\end{figure}

As mentioned in the introduction, Finn and Lizier proposed a partial decomposition based on pointwise mutual information, PPID \cite{PPID}. The authors also introduce a new example called ``Pointwise Unique'', where in any pointwise configuration only one source holds non-zero information about the target. While other existing measures do not identify unique information for the so-called \textsc{PwUnq} distribution, their approach does.
{ 
\renewcommand{\arraystretch}{1.3}  
\begin{table}[H]
\caption{Four of the probability distributions from \cite{PPID}.}
\centering
\begin{tabular}{ccccc|ccccc|ccccc|cccccc}
\toprule
\multicolumn{4}{c}{\textbf{\textsc{PwUnq}}} & & \multicolumn{4}{c}{\textbf{\textsc{RndErr}}} & & \multicolumn{4}{c}{\textbf{\textsc{Unq}}} & & \multicolumn{4}{c}{\textbf{\textsc{And}}}\\
\hline
\emph{X} & \emph{Y} & \emph{Z} & \emph{p} & \textbf{} & \emph{X} & \emph{Y} & \emph{Z} & \emph{p}  & \textbf{} & \emph{X} & \emph{Y} & \emph{Z} & \emph{p} & \textbf{} & \emph{X} & \emph{Y} & \emph{Z} & \emph{p}\\
\hline
0 & 1 & 1  & $\tfrac{1}{4}$ & & 0 & 0 & 0  & $\tfrac{3}{8}$ & & 0 & 0 & 0  & $\tfrac{1}{4}$ & & 0 & 0 & 0  & $\tfrac{1}{4}$\\
1 & 0 & 1  & $\tfrac{1}{4}$ & & 1 & 1 & 1  & $\tfrac{3}{8}$ & & 1 & 1 & 1  & $\tfrac{1}{4}$ & & 1 & 0 & 0  & $\tfrac{1}{4}$\\
0 & 2 & 2  & $\tfrac{1}{4}$ & & 0 & 1 & 0  & $\tfrac{1}{8}$ & & 0 & 1 & 0  & $\tfrac{1}{4}$ & & 0 & 1 & 0  & $\tfrac{1}{4}$\\
2 & 0 & 2  & $\tfrac{1}{4}$ & & 1 & 0 & 1  & $\tfrac{1}{8}$ & & 1 & 0 & 1  & $\tfrac{1}{4}$ & & 1 & 1 & 1  & $\tfrac{1}{4}$\\
\hline
\end{tabular}
\label{tbl:PPID}
\end{table}
} 
As with the previous examples, the probabilities are used to determine the underlying structures. In Figure \ref{PPIDFour} the resulting structures are depicted. Using the proposed definition it is immediately clear from these structures that: \textit{(i)} there is no redundant information in the \textsc{Unq} distribution and the \textsc{And} distribution, and \textit{(ii)} $Y$ does not provide unique information to $Z$ for the \textsc{RndErr} and \textsc{Unq} distributions.

Although the procedure that needs to be followed is now clear, we specifically determine the tensor for the path $\{x\}\{y\}\{z\}$ for the \textsc{PwUnq} distribution using Table \ref{tbl:PPID} and Equation (\ref{eq:ChainTransmissionMI}). Because

\small{
\begin{equation*}
\mathcal{T}\{ x\}\{y \} = 
\begin{pmatrix} 
0 & \tfrac{1}{2} & \tfrac{1}{2}\\
1 & 0 & 0\\
1 & 0 & 0  
\end{pmatrix},\text{ and }
\mathcal{T}\{ y\}\{z \} = 
\begin{pmatrix} 
\tfrac{1}{2} & \tfrac{1}{2}\\
1 & 0\\
0 & 1
\end{pmatrix},\; 
\mathcal{T}\{ x\}\{y \}\{z \} = 
\begin{pmatrix} 
\tfrac{1}{2} & \tfrac{1}{2}\\
\tfrac{1}{2} & \tfrac{1}{2}\\
\tfrac{1}{2} & \tfrac{1}{2}
\end{pmatrix}.
\end{equation*}
}\normalsize

A communication channel with identical rows in the probability transition matrix is incapable of transmitting any information: the redundant information equals \emph{0 bit}. The final PID results for \textsc{PwUnq} distribution and four other distributions are tabulated in Table \ref{tbl:PPIDresult}.
\begin{table}[H]
\caption{PID for \textsc{PwUnq}, \textsc{RndErr}, \textsc{Unq}, \textsc{And}, and the Two Bit Copy, \textsc{Tbc}, from Section \ref{ss:TBC}.}
\centering
\begingroup
\setlength{\tabcolsep}{2pt}
\begin{tabular}{cccc|ccc|ccc|ccc|cc}
\toprule
& \multicolumn{2}{c}{\textbf{\textsc{PwUnq}}} &  & \multicolumn{2}{c}{\textbf{\textsc{RndErr}}} & & \multicolumn{2}{c}{\textbf{\textsc{Unq}}} & & \multicolumn{2}{c}{\textbf{\textsc{And}}}& & \multicolumn{2}{c}{\textbf{\textsc{Tbc}}}\\
\hline
\textbf{Lattice Node} & $I_{\partial}(\langle i \rangle)$ & $I_{\partial}(I_\bigtriangleup)$ & & $I_{\partial}(\langle i \rangle)$ & $I_{\partial}(I_\bigtriangleup)$ & & $I_{\partial}(\langle i \rangle)$ & $I_{\partial}(I_\bigtriangleup)$ & & $I_{\partial}(\langle i \rangle)$ & $I_{\partial}(I_\bigtriangleup)$ & & $I_{\partial}(\langle i \rangle)$ & $I_{\partial}(I_\bigtriangleup)$\\
\hline
$\{ 12 \}$ 			& 0 		& 0   & & 0.811 		& 0		& & 1 			& 0 	& & 0.75 		& 0.19 & & 1 		& 0\\
$\{ 2 \}$ 			& 0.5 	& 0.5 & & -0.811 	& 0 		& & -1 			& 0 	& & -0.25 		& 0.31 & & 0 		& 1\\
$\{ 1 \}$ 			& 0.5 	& 0.5 & & 0  		& 0.811	& & 0 			& 1 	& & -0.25 		& 0.31 & & 0 		& 1\\
$\{ 1 \}$$\{ 2 \}$ 	& 0 		& 0   & & 1  		& 0.189 	& & 1 			& 0	    & & 0.561 		& 0	   & & 1 		& 0\\
\hline
\end{tabular}
\endgroup
\label{tbl:PPIDresult}
\end{table}
From this table it is evident that our proposed PID does give the ``correct'' decomposition for the \textsc{PwUnq} distribution. This implies that some aspects of the pointwise approach are captured by the causal tensors. 
\subsubsection{Comparison with $I_{min}$, $I_{broja}$, and $I_{CCS}$}
In the last section we applied the proposed PID to some examples that strive to have non-negative PID for a system comprising three variables. For the distributions we refer to \cite{Ince}.

\begin{table}[H]
\caption{PID for \textsc{5a}.}
\centering
\begin{tabular}{ccccc}
\toprule
\textbf{Lattice Node} & $I_{\partial}(I_{min})$ & $I_{\partial}(I_{broja})$ 	&  $I_{\partial}(I_{CCS})$ & $I_{\partial}(I_\bigtriangleup)$\\
\hline
$\{ 12 \}$ 			& 0.3333 	& 0  		& 0.1383 		& 0.1383  		\\
$\{ 2 \}$ 			& 0.3333 	& 0.6666  	& 0.5283 		& 0.5283 		\\
$\{ 1 \}$ 			& 0.3333 	& 0.6666 	& 0.5283 		& 0.5283  		\\
$\{ 1 \}$$\{ 2 \}$ 	& 0.5850 	& 0.2516  	& 0.3900 		& 0.3900   		\\
\hline
\end{tabular}
\label{tbl:5A}
\end{table}

\begin{table}[H]
\caption{PID for \textsc{5b}.}
\centering
\begin{tabular}{ccccc}
\toprule
\textbf{Lattice Node} & $I_{\partial}(I_{min})$ & $I_{\partial}(I_{broja})$ 	&  $I_{\partial}(I_{CCS})$ & $I_{\partial}(I_\bigtriangleup)$\\
\hline
$\{ 12 \}$ 			& 0.5 	& 0  	& 0 		& 0  		\\
$\{ 2 \}$ 			& 0.5 	& 1  	& 1 		& 1 			\\
$\{ 1 \}$ 			& 0 		& 0.5 	& 0.5 	& 0.5  		\\
$\{ 1 \}$$\{ 2 \}$ 	& 0.5 	& 0  	& 0 		& 0   		\\
\hline
\end{tabular}
\label{tbl:5B}
\end{table}

\begin{table}[H]
\caption{PID for \textsc{5c}.}
\centering
\begin{tabular}{ccccc}
\toprule
\textbf{Lattice Node} & $I_{\partial}(I_{min})$ & $I_{\partial}(I_{broja})$ 	&  $I_{\partial}(I_{CCS})$ & $I_{\partial}(I_\bigtriangleup)$\\
\hline
$\{ 12 \}$ 			& 0.67 	& 0.67  	& 0.67 	& 0.67  		\\
$\{ 2 \}$ 			& 0.25 	& 0.25  	& 0.25 	& 0.25 		\\
$\{ 1 \}$ 			& 0 		& 0 		& 0	 	& 0  		\\
$\{ 1 \}$$\{ 2 \}$ 	& 0  	& 0  	& 0 		& 0   		\\
\hline
\end{tabular}
\label{tbl:5C}
\end{table}

\begin{table}[H]
\caption{PID for \textsc{ReducedOr}.}
\centering
\begin{tabular}{ccccc}
\toprule
\textbf{Lattice Node} & $I_{\partial}(I_{min})$ & $I_{\partial}(I_{broja})$ 	&  $I_{\partial}(I_{CCS})$ & $I_{\partial}(I_\bigtriangleup)$\\
\hline
$\{ 12 \}$ 			& 0.69 	& 0.69  	& 0.38 		& 0.40  		\\
$\{ 2 \}$ 			& 0 		& 0  	& 0.31 		& 0.29 		\\
$\{ 1 \}$ 			& 0 		& 0  	& 0.31 		& 0.29   	\\
$\{ 1 \}$$\{ 2 \}$ 	& 0.31	& 0.31 	& 0 			& 0.02  		\\
\hline
\end{tabular}
\label{tbl:ReducedOr}
\end{table}

\begin{table}[H]
\caption{PID for \textsc{xor}.}
\centering
\begin{tabular}{ccccc}
\toprule
\textbf{Lattice Node} & $I_{\partial}(I_{min})$ & $I_{\partial}(I_{broja})$ 	&  $I_{\partial}(I_{CCS})$ & $I_{\partial}(I_\bigtriangleup)$\\
\hline
$\{ 12 \}$ 			& 1 & 1  & 1 & 1  	\\
$\{ 2 \}$ 			& 0 & 0  & 0 & 0 	\\
$\{ 1 \}$ 			& 0 & 0  & 0 & 0   	\\
$\{ 1 \}$$\{ 2 \}$ 	& 0 & 0  & 0 & 0   	\\
\hline
\end{tabular}
\label{tbl:XOR}
\end{table}

An interesting difference between other proposed methods and our proposed method is that there is no redundancy in the \textsc{And/Or} distributions. 
\begin{table}[H]
\caption{PID for \textsc{And/Or}.}
\centering
\begin{tabular}{ccccc}
\toprule
\textbf{Lattice Node} & $I_{\partial}(I_{min})$ & $I_{\partial}(I_{broja})$ 	&  $I_{\partial}(I_{CCS})$ & $I_{\partial}(I_\bigtriangleup)$\\
\hline
$\{ 12 \}$ 			& 0.5 	& 0.5  	& 0.29		& 0.19  		\\
$\{ 2 \}$ 			& 0 		& 0  	& 0.21 		& 0.31 		\\
$\{ 1 \}$ 			& 0 		& 0  	& 0.21 		& 0.31   	\\
$\{ 1 \}$$\{ 2 \}$ 	& 0.31 	& 0.31  	& 0.10 		& 0   		\\
\hline
\end{tabular}
\label{tbl:AND}
\end{table}

\begin{table}[H]
\caption{PID for \textsc{Sum}.}
\centering
\begin{tabular}{ccccc}
\toprule
\textbf{Lattice Node} & $I_{\partial}(I_{min})$ & $I_{\partial}(I_{broja})$ 	&  $I_{\partial}(I_{CCS})$ & $I_{\partial}(I_\bigtriangleup)$\\
\hline
$\{ 12 \}$ 			& 1 		& 1  	& 0.5 		& 0.5  		\\
$\{ 2 \}$ 			& 0 		& 0  	& 0.5 		& 0.5 		\\
$\{ 1 \}$ 			& 0 		& 0  	& 0.5 		& 0.5   		\\
$\{ 1 \}$$\{ 2 \}$ 	& 0.5 	& 0.5  	& 0 			& 0   		\\
\hline
\end{tabular}
\label{tbl:SUM}
\end{table}
From these examples, it is clear that our proposed PID closely resembles the PID proposed by Ince \cite{Ince}. This should not come as a surprise because the pointwise approach suggested by Ince expresses associations in terms of mutual information a posteriori.\\

In this article, we have shown that a partial information decomposition comprising non-negative unique and redundant contributions follows naturally from the framework of causal tensors. Because we introduced no new information theoretical measures, it is our contention that a partial information decomposition is possible within the framework of ``classical'' Shannon information theory. A partial information decomposition in terms of ``average measures'' is problematic when no exact expressions for indirect paths can be determined. It reduces to a rather straightforward exercise when this is possible, for example, within the framework of causal tensors. What needs to be investigated is that the proposed approach does not find any redundancy in the \textsc{And/Or} case, while it is commonly accepted that there should be redundancy. It was furthermore demonstrated that negative information components could be used for diagnostic purposes.

Future research will be focussed on expanding the proposed PID to systems comprising over four variables using matrix operations. The goal is to investigate the usefulness of the proposed PID for diagnosing a complex electromechanical system comprising over several hundred components.\\

\funding{This research received no external funding.}
\acknowledgments{I would like to thank Ryan James for providing examples to compare against.}
\conflictsofinterest{The authors declare no conflict of interest.} 

\appendixtitles{yes}
\appendix
\section{Proof of Lemma \ref{col:opposite_traverse}} \label{App:TraversePath} 
\begin{proof}
	Let 
	
\begin{equation} \label{eq:MIpath}
	I\{ x\}\{y\}\{z\} = \sum_{i,k} p^{ik} \log_2\left[ \frac{\sum_j A^j_i B^{k_1}_{j}}{p^{k}} \right].
\end{equation}

Because $A^j_i\!=\!A^{\ddagger i}_j p^j\big/ p^i$, and $B^{k_1}_{j}\!=\!B^{\ddagger j}_{k_1} p^{k_1}\big/ p^j$, Equation (\ref{eq:MIpath}) can be rewritten as
\begin{equation*}
	I\{ x\}\{y\}\{z\} = \sum_{i,k} p^{ik} \log_2\left[ \frac{\sum_j B^{\ddagger j}_{k} A^{\ddagger i}_j}{p^{i}} \right],
\end{equation*}

i.e., $I\{ x\}\{y\}\{z\} = I\{z\}\{y\}\{ x\}$.\\
\end{proof}
\section{Proof of Theorem \ref{Thm:SymmetryOfRedundancy}}\label{App:SymmetryOfRedundancy}
For the proof we make use of Lemma \ref{col:opposite_traverse} and Definition \ref{def:Redundancy}. These are used to create expressions for $\mathcal{R}(X,Y;Z)$, $\mathcal{R}(Z,Y;X)$, and $\mathcal{R}(Z,X;Y)$. 
\begin{proof}
Using Definition \ref{def:Redundancy}, we get

\begin{align*}	
	\mathcal{R}(X,Y;Z) &= \min \left[ I\{x\}\{y\}\{z\},I\{y\}\{x\}\{z\}  \right],\\
	\mathcal{R}(Z,Y;X) &= \min \left[ I\{z\}\{y\}\{x\},I\{y\}\{z\}\{x\}  \right],\\
	\mathcal{R}(Z,X;Y) &= \min \left[ I\{z\}\{x\}\{y\},I\{x\}\{z\}\{y\}  \right].
\end{align*}

Applying Lemma \ref{col:opposite_traverse} and using $\alpha\!=\!I\{x\}\{y\}\{z\}$, $\beta\!=\!I\{y\}\{x\}\{z\}$, $\gamma\!=\!I\{y\}\{z\}\{x\}$, $R_1\!=\!\mathcal{R}(X,Y;Z)$, $R_2\!=\!\mathcal{R}(Z,Y;X)$, and $R_3\!=\!\mathcal{R}(Z,X;Y)$, this set of redundancies can be simplified to:

\begin{align} \label{eq:SetOfRedundancies}	
	R_1 &= \min \left[ \alpha,\beta  \right],\\
	R_2 &= \min \left[ \alpha,\gamma \right],\\
	R_3 &= \min \left[ \beta ,\gamma  \right].
\end{align}

We will now investigate all possible combinations of outputs. Because we are looking for the situation where $R_1$ is larger than or equal either $R_2$ or $R_3$, we stop once such a situation is found. The other non relevant entries are set to $N.A.$, i.e., not applicable.

\begin{table}[H]
\caption{All possible combinations for the redundancies from Equation (\ref{eq:SetOfRedundancies}).}
\centering
\begin{tabular}{cccl}
\toprule
$R_1$ & $R_2$ 	&  $R_3$ & $Conclusion$\\
\hline
$\alpha$ 	& $\alpha$  		& $N.A.$		& $R_1 \geq R_2$, because $R_1 = R_2$  		\\
$\alpha$ 	& $\gamma$  		& $N.A.$	 	& $R_1 \geq R_2$, because for $R_2$, $\gamma \leq \alpha$ 		\\
$\beta$ 	& $\alpha$ 		& $\beta$ 	& $R_1 \geq R_3$, because $R_1 = R_3$  		\\
$\beta$ 	& $\alpha$ 		& $\gamma$	& $R_1 \geq R_3$, because for $R_3$, $\gamma \leq \beta$  		\\
$\beta$ 	& $\gamma$ 		& $\beta$ 	& $R_1 \geq R_3$, because $R_1 = R_3$  	  	\\
$\beta$ 	& $\gamma$ 		& $\gamma$ 	& $R_1 \geq R_3$, because for $R_3$, $\gamma \leq \beta$   		\\
\hline
\end{tabular}
\label{tbl:SetOfRedundancies}
\end{table}
From this table it follows that
\begin{equation}
	\mathcal{R}(X,Y;Z) \geq \mathcal{R}(Z,Y;X) \text{ or } \mathcal{R}(X,Y;Z) \geq \mathcal{R}(Z,X;Y).
\end{equation}
\end{proof}
\section{Proof of Identity Property (Equation \ref{eq:Identity})}\label{App:Identity}
\begin{proof}
Per definition

\begin{equation*}
\mathcal{R}(X,Y;XY) = \min \left[ I\{ x\}\{y\}\{xy \}\!, I\{ y\}\{x\}\{xy \} \right]. 
\end{equation*}

Assume that, in index notation, $p^i$ equals the pmf of $X$, $p^j$ equals the pmf of $Y$, and $p^{ij}$ equals the pmf of $XY$. Assume furthermore that $A^j_i$ are the tensor elements for the path $\{ x\}\{y \}$. Per definition $A^{\ddagger i}_j$ represent the tensor elements for the path $\{ y\}\{x \}$. Let tensor elements for the path $\{ x\}\{xy \}$ equal $C^{ij}_{i’}$, and the tensor elements for the path $\{ y\}\{xy \}$ equal $B^{ij}_{j’}$. Using these tensors, the redundancy equals

\begin{equation*}
 \mathcal{R}(X,Y;XY) = \min \left[ \sum_{i,j} p^{ij}\log_2 \left[ \frac{\sum_{j’} A^{j’}_{i}B^{ij}_{j’}}{p^{ij}} \right], \sum_{i,j} p^{ij}\log_2 \left[ \frac{\sum_{i’} A^{\ddagger i’}_jC^{ij}_{i’}}{p^{ij}} \right] \right].
\end{equation*}

The tensor elements $B^{ij}_{j’}$ are related to $A^{\ddagger i}_j$. It follows immediately that $B^{ij}_{j’=j}= A^{\ddagger i}_j$ and $B^{ij}_{j’ \neq j}=0$, Likewise, $C^{ij}_{i’=i}= A^{j}_i$ and $C^{ij}_{i’ \neq i}=0$. We can rewrite the joint probabilities as $p^{ij} = p^i A^j_i$, and as $p^{ij} = p^j A^{\ddagger i}_j$. Using these expressions, we rewrite the equation for the redundancy as

\begin{equation*}
 \mathcal{R}(X,Y;XY) = \min \left[ \sum_{i,j} p^{ij}\log_2 \left[ \frac{A^{\ddagger i}_j}{p^{i}} \right], \sum_{i,j} p^{ij}\log_2 \left[ \frac{A^j_i}{p^{j}} \right] \right].
\end{equation*}

Because both sums represent $I(X;Y)$, the mutual information between $X$ and $Y$ \cite{DSIG}, the proposed redundancy measure satisfies the identity property.
\end{proof}
\section{Proof of Left Monotonicity (see Equation (\ref{eq:LeftMono}))}\label{App:LeftMono}

For the proof of the left monotonicity the Redundancy lattice of Figure \ref{RLattice}b, Theorem \ref{Thm:SymmetryOfRedundancy} and the symmetry axiom are used.
\begin{proof}
	We start with the redundancy $\mathcal{R}(X,Y;Z_1Z_2)$. Using Theorem \ref{Thm:SymmetryOfRedundancy} we get

\begin{equation} \label{eq:RedundancyPath}
\mathcal{R}(X,Y;Z_1Z_2) \geq     
	\begin{cases}
            \mathcal{R}(Z_1Z_2,Y;X),\\
            \;\;\;\;\;\;\textsc{Or}\\
            \mathcal{R}(Z_1Z_2,X;Y).
    \end{cases}
\end{equation}

Replacing $Z_1$ with $1$, $Z_2$ with $2$, $Y$ with $3$, or $X$ with $3$, it follows from Figure \ref{RLattice}b that the inequality of Equation (\ref{eq:RedundancyPath}) gives rise to the inequality

\begin{equation} \label{eq:RedundancyPath2}
\mathcal{R}(X,Y;Z_1Z_2) \geq 
	\begin{cases}
            \mathcal{R}(Z_1,Y;X),\\
            \;\;\;\;\;\;\textsc{Or}\\
            \mathcal{R}(Z_1,X;Y).
    \end{cases}	
\end{equation}

We now apply Theorem \ref{Thm:SymmetryOfRedundancy} a second time. Either

\begin{equation} \label{eq:RedundancyPath3a}
\mathcal{R}(X,Y;Z_1Z_2) \geq \mathcal{R}(Z_1,Y;X) \geq
	\begin{cases}
            \mathcal{R}(X,Y;Z_1),\\
            \;\;\;\;\;\;\textsc{Or}\\
            \mathcal{R}(X,Z_1;Y),
    \end{cases}	
\end{equation}

\textsc{Or} 

\begin{equation} \label{eq:RedundancyPath3b}
\mathcal{R}(X,Y;Z_1Z_2) \geq \mathcal{R}(Z_1,X;Y) \geq
	\begin{cases}
            \mathcal{R}(Y,X;Z_1),\\
            \;\;\;\;\;\;\textsc{Or}\\
            \mathcal{R}(Y,Z_1;X).
    \end{cases}	
\end{equation}

Using the symmetry axiom and writing $\mathcal{R}(Z_1,Y;X)$ as $\mathcal{R}_1$, and $\mathcal{R}(Z_1,X;Y)$ as $\mathcal{R}_2$, we get either

\begin{equation} \label{eq:RedundancyPath4a}
\mathcal{R}(X,Y;Z_1Z_2) \geq \mathcal{R}_1 \geq
	\begin{cases}
            \mathcal{R}(X,Y;Z_1),\\
            \;\;\;\;\;\;\textsc{Or}\\
            \mathcal{R}_2,
    \end{cases}	
\end{equation}

\textsc{Or} 

\begin{equation} \label{eq:RedundancyPath4b}
\mathcal{R}(X,Y;Z_1Z_2) \geq \mathcal{R}_2 \geq
	\begin{cases}
            \mathcal{R}(X,Y;Z_1),\\
            \;\;\;\;\;\;\textsc{Or}\\
            \mathcal{R}_1.
    \end{cases}	
\end{equation}

We assume that $\mathcal{R}_2 \neq \mathcal{R}_1$. Equality only occurs when the relation between $X$ and $Y$ is noiseless. If in Equation (\ref{eq:RedundancyPath4a}) $\mathcal{R}_1 \geq \mathcal{R}_2$, then it follows from Equation (\ref{eq:RedundancyPath4b}) that $\mathcal{R}_2 \geq \mathcal{R}(X,Y;Z_1)$. If on the other hand $\mathcal{R}_2 \geq \mathcal{R}_1$ in Equation (\ref{eq:RedundancyPath4b}), then it follows from Equation (\ref{eq:RedundancyPath4a}) that $\mathcal{R}_1 \geq \mathcal{R}(X,Y;Z_1)$. In other words,

\begin{equation} \label{Redundancy2Lattice}
	\mathcal{R}(X,Y;Z_1Z_2) \geq \mathcal{R}(X,Y;Z_1). 
\end{equation}
\\
\end{proof}

\section{Demonstration of the non-existence of a path between $X$ and $Y$ in case $\tilde{Z}$ is observed Section \ref{Section:Really}}\label{App:UnObserved}
See Figure \ref{Unobserved} for reference. Using table \ref{tbl:NegativeSynergy}, the reader can verify that $\mathcal{T}\{ x\}\{y \} \!=\! \mathcal{T}\{ x\}\{\tilde{z} \} \cdot \mathcal{T}\{ \tilde{z}\}\{y \}$, because 

\small{
\begin{equation*}
\mathcal{T}\{ x\}\{y \} = 
\begin{pmatrix} 
\tfrac{1}{2} & \tfrac{1}{2}\\
0 & 1  
\end{pmatrix},\;
\mathcal{T}\{ x\}\{\tilde{z} \} = 
\begin{pmatrix} 
\tfrac{1}{2} & \tfrac{1}{2} & 0 & 0\\
0 & 0 &\tfrac{1}{2} & \tfrac{1}{2}
\end{pmatrix},\text{ and }
\mathcal{T}\{ \tilde{z}\}\{y \} = 
\begin{pmatrix} 
1 & 0\\
0 & 1\\  
0 & 1\\
0 & 1 
\end{pmatrix}. 
\end{equation*}
}\normalsize

This implies that the $path \{ x\}\{y \}$ can be fully explained by the indirect $path \{ x\}\{ y\}\{\tilde{z} \}$. However, if the common cause is unobserved (see Figure \ref{Unobserved}b), there is no reason to state that the path is nonexistent. Using Table \ref{tbl:NegativeSynergy} we find that

\small{
\begin{equation*}
\mathcal{T}\{ x\}\{z \} = 
\begin{pmatrix}
1 & 0\\
\tfrac{1}{2} & \tfrac{1}{2}
\end{pmatrix},
\mathcal{T}\{ z\}\{y \} = 
\begin{pmatrix} 
\tfrac{1}{3} & \tfrac{2}{3}\\
0 & 1
\end{pmatrix}, \text{ and }
\mathcal{T}\{ x\}\{z \} \cdot \mathcal{T}\{ z\}\{y \} = 
\begin{pmatrix}
\tfrac{1}{3} & \tfrac{2}{3}\\
\tfrac{1}{6} & \tfrac{5}{6}\\
\end{pmatrix},
\end{equation*}
}\normalsize

i.e., $\mathcal{T}\{ x\}\{y \} \!\neq\! \mathcal{T}\{ x\}\{z \} \cdot \mathcal{T}\{ z\}\{y \}$. In a similar fashion it follows that $\mathcal{T}\{ y\}\{x \} \!\neq\! \mathcal{T}\{ y\}\{z \} \cdot \mathcal{T}\{ z\}\{x \}$. This means that the structure is a fully connected system and that redundant information is contributed via the paths $\{x\}\!\{y\}\!\{z\}$ and $\{y\}\!\{x\}\!\{z\}$. 


\end{document}